\documentclass[twocolumn,amsmath,amssymb,showpacs,superscriptaddress]{revtex4}

\usepackage{graphicx}
\usepackage{dcolumn}
\usepackage{bm}

\def\pz{\pi^0}
\def\ee{e^+e^-}
\def\ll{l^+l^-}
\def\mm{\mu^+\mu^-}
\def\g{\gamma}
\def\ep{\epsilon}

\begin{document}

\preprint{APS/123-QED}

\title{Final Results from the KTeV Experiment on the Decay $K_L\to\pz\g\g$}

%

\newcommand{\UAz}{University of Arizona, Tucson, Arizona 85721}
\newcommand{\UCLA}{University of California at Los Angeles, Los Angeles,
                    California 90095} 
\newcommand{\Campinas}{Universidade Estadual de Campinas, Campinas, 
                       Brazil 13083-970}
\newcommand{\EFI}{The Enrico Fermi Institute, The University of Chicago, 
                  Chicago, Illinois 60637}
\newcommand{\UB}{University of Colorado, Boulder, Colorado 80309}
\newcommand{\ELM}{Elmhurst College, Elmhurst, Illinois 60126}
\newcommand{\FNAL}{Fermi National Accelerator Laboratory, 
                   Batavia, Illinois 60510}
\newcommand{\Osaka}{Osaka University, Toyonaka, Osaka 560-0043 Japan} 
\newcommand{\Rice}{Rice University, Houston, Texas 77005}
\newcommand{\SaoPaolo}{Universidade de S\~ao Paulo, S\~ao Paulo, Brazil 05315-970}
\newcommand{\UVa}{The Department of Physics and Institute of Nuclear and 
                  Particle Physics, University of Virginia, 
                  Charlottesville, Virginia 22901}
\newcommand{\UW}{University of Wisconsin, Madison, Wisconsin 53706}

\affiliation{\UAz}
\affiliation{\UCLA}
\affiliation{\Campinas}
\affiliation{\EFI}
\affiliation{\UB}
\affiliation{\ELM}
\affiliation{\FNAL}
\affiliation{\Osaka}
\affiliation{\Rice}
\affiliation{\SaoPaolo}
\affiliation{\UVa}
\affiliation{\UW}

\author{E.~Abouzaid}      \affiliation{\EFI}
\author{M.~Arenton}       \affiliation{\UVa}
\author{A.R.~Barker}      \altaffiliation[Deceased.]{ } \affiliation{\UB}
\author{L.~Bellantoni}    \affiliation{\FNAL}
\author{E.~Blucher}       \affiliation{\EFI}
\author{G.J.~Bock}        \affiliation{\FNAL}
\author{E.~Cheu}          \affiliation{\UAz}
\author{R.~Coleman}       \affiliation{\FNAL}
\author{M.D.~Corcoran}    \affiliation{\Rice}
\author{B.~Cox}           \affiliation{\UVa}
\author{A.R.~Erwin}       \affiliation{\UW}
\author{C.O.~Escobar}     \affiliation{\Campinas}  
\author{A.~Glazov}        \affiliation{\EFI}
\author{A.~Golossanov}    \affiliation{\UVa} 
\author{R.A.~Gomes}       \affiliation{\Campinas}
\author{P. Gouffon}       \affiliation{\SaoPaolo}
\author{Y.B.~Hsiung}      \affiliation{\FNAL}
\author{D.A.~Jensen}      \affiliation{\FNAL}
\author{R.~Kessler}       \affiliation{\EFI}
\author{K.~Kotera}        \affiliation{\Osaka}
\author{A.~Ledovskoy}     \affiliation{\UVa}
\author{P.L.~McBride}     \affiliation{\FNAL}

\author{E.~Monnier}
   \altaffiliation[Permanent address ]{C.P.P. Marseille/C.N.R.S., France}
   \affiliation{\EFI}  

\author{H.~Nguyen}       \affiliation{\FNAL}
\author{R.~Niclasen}     \affiliation{\UB}
\author{D.G.~Phillips~II} \affiliation{\UVa}
\author{E.J.~Ramberg}    \affiliation{\FNAL}
\author{R.E.~Ray}        \affiliation{\FNAL}
\author{M.~Ronquest}     \affiliation{\UVa}
\author{E.~Santos}       \affiliation{\SaoPaolo}
\author{W.~Slater}       \affiliation{\UCLA}
\author{D.~Smith}        \affiliation{\UVa}
\author{N.~Solomey}      \affiliation{\EFI}
\author{E.C.~Swallow}    \affiliation{\EFI}\affiliation{\ELM}
\author{P.A.~Toale}      \affiliation{\UB}
\author{R.~Tschirhart}   \affiliation{\FNAL}
\author{Y.W.~Wah}        \affiliation{\EFI}
\author{J.~Wang}         \affiliation{\UAz}
\author{H.B.~White}      \affiliation{\FNAL}
\author{J.~Whitmore}     \affiliation{\FNAL}
\author{M.~J.~Wilking}      \affiliation{\UB}
\author{B.~Winstein}     \affiliation{\EFI}
\author{R.~Winston}      \affiliation{\EFI}
\author{E.T.~Worcester}  \affiliation{\EFI}
\author{T.~Yamanaka}     \affiliation{\Osaka}
\author{E.~D.~Zimmerman} \affiliation{\UB}
\author{R.F.~Zukanovich} \affiliation{\SaoPaolo}


\date{\today}

\begin{abstract}
We report on a new measurement of the branching ratio B($K_L\to\pz\g\g$)
 using the KTeV detector.
We reconstruct 1982 events with an estimated background of 608, 
that results in
B($K_L\to\pz\g\g$) = $(1.29 \pm 0.03_{stat} \pm 0.05_{syst})
\times 10^{-6}$. We also measure the parameter, $a_V$, which 
characterizes the strength of vector meson exchange terms in
this decay. We find $a_V = -0.31 \pm 0.05_{stat} \pm 0.07_{syst}$.
These results utilize the full KTeV data set collected
from 1997 to 2000 and supersede earlier KTeV measurements of the
branching ratio and $a_V$.
\end{abstract}

\maketitle

\section{\label{sec:intro}Introduction}

The decay $K_L\to\pz\g\g$ provides important checks of
low-energy theories of strange meson decays. In Chiral Perturbation
Theory (ChPT) the branching ratio for this decay can be determined
with no free parameters up to $O(p^4)$. However, the first measurements
of the branching ratio for $K_L\to\pz\g\g$
\cite{ref:na31a, ref:e731, ref:na31b}
were approximately three times larger than
the predicted $O(p^4)$ branching ratio of 
$0.68 \times 10^{-6}$\cite{ref:ecker}.
Extending the theory to $O(p^6)$ and including vector meson
exchange terms raise the branching ratio prediction
to be consistent with the measured values\cite{ref:dambrosio, ref:gabbiani}.
The vector meson contributions can be parametrized by an effective
coupling constant $a_V$. Non perturbative calculations for the
$K_L\to\pz\g\g$ rate have also been performed\cite{ref:truong}.

The $K_L\to\pz\g\g$ decay is important also because of its implications for the
related decay $K_L\to\pz\ll$, where $\ll$ can be either $\ee$ or $\mm$. 
Currently, the best limits for these decays are
B($K_L\to\pz\ee$) $< 2.8\times 10^{-10}$\cite{ref:pi0ee} and
B($K_L\to\pz\mm$) $< 3.8\times 10^{-10}$\cite{ref:pi0mm}, 
both at the 90\% confidence level. The expected
branching ratios are approximately $1-43\times 10^{-11}$\cite{ref:buchalla,ref:mescia}.
There are three contributions to the $K_L\to\pz\ll$ decay, classified in terms
of their CP symmetry; one conserves CP symmetry, one violates
it indirectly and one directly.
The direct CP violating amplitude is of interest within
the Standard Model but also can show signs of 
new physics \cite{ref:mescia}, leading to an enhancement
of the $K_L\to\ll$ rate. In order to determine
the direct CP violating terms, one must first determine the
other two amplitudes. 
The indirect CP violating amplitude 
can be determined from the decay $K_S\to\pz\ll$, and the
NA48 experiment has measured 
B($K_S\to\pz\ee$) $ = 5.8^{+2.9}_{-2.4}\times 10^{-9}$.\cite{ref:ksp0ee}.
Because the $K_L\to\pz\ll$ decay can proceed via a
CP conserving two-photon exchange, 
the CP conserving terms can be probed using $K_L\to\pz\g\g$.
A precise measurement of
the parameter $a_V$ can be used to determine the size of the
CP conserving amplitude in $K_L\to\pz\ll$.

There have been a number of previous measurements of $K_L\to\pz\g\g$ from
the E731, NA31, NA48 and KTeV experiments.\cite{ref:na31a, ref:e731,
ref:na31b, ref:ktev, ref:NA48}.
The two most recent measurements are the NA48 result of 
$(1.36\pm 0.03_{stat}\pm 0.03_{syst} \pm 0.03_{norm})\times 10^{-6}$ 
and the KTeV result of
$(1.68\pm 0.07_{stat}\pm0.08_{syst})\times 10^{-6}$. Both of these
results are significantly more precise than the E731 and NA31 results. 
However, the
NA48 and KTeV results differ by nearly three standard deviations. 
The measurement discussed here supersedes the
previous KTeV result and reconciles the difference between
these two results.

\section{The KTeV Detector}

Data used in this analysis were collected during three running
periods in 1996, 1997 and 1999 using the KTeV detector at Fermilab.
Because of the similar topology between the $K_L\to\pz\g\g$ decays
and the $K_L\to\pz\pz$ decays used to measure $\ep'/\ep$
\cite{ref:epp97},
we recorded
the $K_L\to\pz\g\g$ events during the same collection period used
for the KTeV $\ep'/\ep$ measurement.

\begin{figure}
\includegraphics[width=8.5cm]{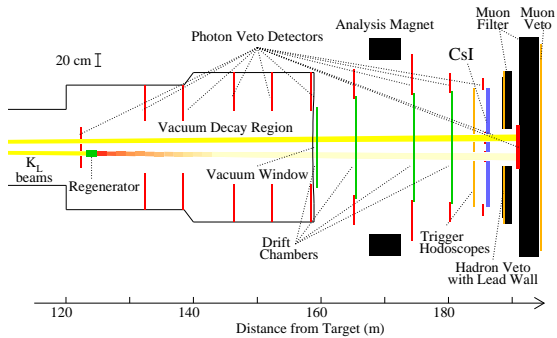}
\caption{\label{fig:ktevdet} Schematic of the KTeV detector.
}
\end{figure}

The KTeV experiment\cite{ref:detector} is a fixed-target experiment built
to study decays of neutral kaons. A schematic of the detector is
shown in Figure~\ref{fig:ktevdet}.
Two neutral kaon beams were produced through interactions of 
800 GeV/$c$ protons in a 30 cm long beryllium oxide target. 
The resulting neutral
particles passed through a series of collimators and absorbers to produce
two nearly parallel beams. 
Charged particles were removed from the beams by sweeping
magnets located downstream of the collimators.
A vacuum decay volume extended from 94 to 159 meters downstream
of the target, and was far enough away from the
target that the vast majority of the $K_S$ component had decayed
away. 
An active regenerator was located within the vacuum region, approximately 123 meters
downstream of the target.
This regenerator alternated between the two
neutral beams to generate a $K_S$ component in one of the beams.
The beam that coincided with the regenerator was called the regenerator
beam, while the other beam was denoted the vacuum beam.
For this analysis, we only considered decays from the
vacuum beam. 
To reject photons, primarily from decays
of $K_L\to\pz\pz\pz$, the 
decay volume was surrounded by photon veto detectors,
that rejected photons produced at angles
greater than 100 milliradians with laboratory energies greater
than 100 MeV. A kevlar and mylar vacuum window with a radiation
length of 0.14\% covered the downstream end of the
vacuum decay region.

\begin{figure}
\includegraphics[width=8.5cm]{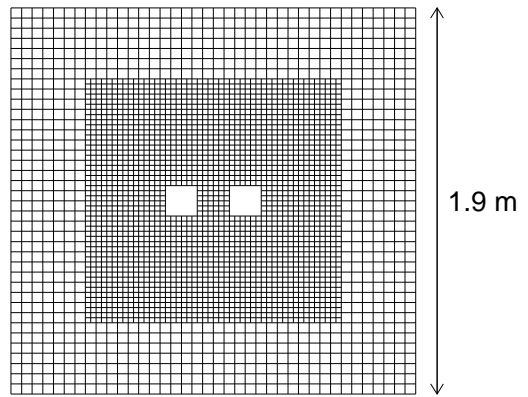}
\caption{\label{fig:csi} Transverse view of the KTeV CsI calorimeter.
The smaller blocks are located in the central region with the larger blocks
located in the outer region.
}
\end{figure}

The most critical detector element in this analysis was
the pure CsI electromagnetic calorimeter\cite{ref:epp97}.
The CsI calorimeter, shown in Figure~\ref{fig:csi},
was composed of 3100 blocks in a 1.9 m by 1.9 m array
with a depth of 50 cm corresponding to 
27 radiation lengths. Two 15 cm by 15 cm holes were located
near the center of the array for the passage of the neutral beams.
For photons with energies between 2 and 60 GeV,
the calorimeter energy resolution was below
1\% and the nonlinearity was less than 0.5\% per 100 GeV. 
The position resolution
of the calorimeter was approximately 1 mm.

Two levels of hardware triggers were used in the KTeV experiment.
For the $K_L\to\pz\g\g$ events, the first level trigger required
the event to 
deposit more than approximately 25 GeV in the CsI 
calorimeter with less than 100 MeV in any of the photon vetoes.  
The second level trigger utilized a hardware cluster 
processor that counted the number of separate clusters in the
CsI  calorimeter\cite{hcc}. Each cluster had to have an energy
greater than approximately 1.0 GeV and the total number
of clusters in the CsI calorimeter was required to be 
equal to four.

Events that satisfied the hardware triggers were also required to 
satisfy an online software filter.
This filter required 
that one of the six possible combinations of two photons reconstructed
near the $\pz$ mass. In addition, the filter required that the decay
vertex reconstructed  upstream of 155 m  for the 1996-1997 data and 
140 m for the 1999 data. This requirement was tightened for the 1999 data
to help reduce the trigger rate for the $K_L\to\pz\g\g$ sample.
These trigger requirements also selected $K_L\to\pz\pz$ 
events which were used as a normalization mode for calculating
the $K_L\to\pz\g\g$ branching ratio.

\section{Event Reconstruction}
\label{sec:recon}
The $K_L\to\pz\g\g$ final state consists of four photons
with no other activity in the detector. In this analysis
we require that all events have exactly four clusters in the
CsI calorimeter and that the energy of each cluster is
greater than 2.0 GeV. To reduce contamination from
events originating from the regenerator beam, 
the center of energy (Eq.~9 of \cite{ref:epp97}) is required to
be within the CsI beam hole corresponding to the vacuum beam.

In the decay $K_L\to\pz\g\g$ the positions and energies of the four photons
do not provide enough constraints
to determine both the decay position and the invariant mass of the system.
Therefore, we assume that the four-photon invariant mass is equal to
the kaon mass, and
reconstruct the decay vertex position ($z$) from the 
calorimeter information. For a $\pz$ decaying
into two photons, 
one can determine the two-photon mass  using the following relation:
\begin{equation}
    m_{12} \approx \frac{\sqrt{E_1E_2}r_{12}}{\Delta z_{CsI}}
\end{equation}
where $E_1$ and $E_2$ are the energies of the two photons,
$r_{12}$ is the distance between the two photons at the CsI
calorimeter, and $\Delta z_{CsI}$ is the distance between the decay
vertex and the CsI calorimeter. Using the position of
the reconstructed decay vertex, we determine the two-photon
mass for each of the six possible combinations and choose
the combination with the reconstructed mass closest to
the known $\pz$ mass. If the closest mass combination differs by
more than 3 MeV/$c^2$ from the known $\pz$ mass, we reject the event.
The total energy of the
kaon system, determined from summing the energies of
the four clusters, is required to be between 40 and 160 GeV.
After these requirements the data sample is dominated by backgrounds
from $K_L\to\pz\pz\pz$ and $K_L\to\pz\pz$ decays. Additional
cuts described in Section~\ref{sec:bkg} are used to reduce these backgrounds.

\section{Monte Carlo Simulation}
A detailed Monte Carlo simulation was used to estimate the 
detector acceptance and the background level in our final sample.
Our Monte Carlo simulates the kaon production at the target and
propagates the kaon amplitude through the detector. The kaon then
decays according to the appropriate decay mode, and the resulting
daughter particles are traced through the KTeV detector. The
interaction of the decay products with the detector 
is simulated and the detector response is then digitized. 

Details of the simulation for all detector components are given
in \cite{ref:epp97}; here we focus on the simulation of the
CsI calorimeter.
To simulate the response of the CsI calorimeter, we used
a library of photons generated using GEANT simulations\cite{ref:GEANT}.
The library contained information deposited into a $13\times 13$
array of CsI crystals. The wrapping and shims separating each
crystal was included in these simulations.
This library was binned as a function of the energy and position
of the incident photon.
We stored the energy depositions for each crystal
in 10 longitudinal bins to 
include the effects of nonlinear response along the 
length of the crystal.

During the course of our studies we found that the GEANT-based
shower library was not adequate for describing the transverse
distribution of the energy in a electromagnetic shower. As noted
below, we make use of this transverse shape to help reduce
backgrounds from $K_L\to\pz\pz\pz$ decays. To better simulate the
shower shapes, we also implemented a data-based shower library.
These showers were extracted from $K_L\to\pz\pz$ events taken
during special, low-intensity runs to reduce the effects of
accidental activity in the CsI calorimeter. The data-based
shower library was also binned as a function of the incident
photon energy and the incident position. 

\begin{figure}
\includegraphics[width=8.5cm]{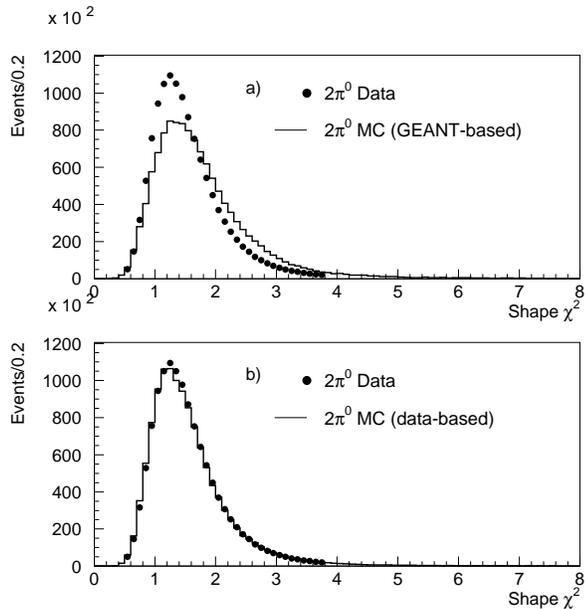}
\caption{\label{fig:fusecomp} The photon shape  $\chi^2$ variable.
a) The photon shape $\chi^2$ variable for $K_L\to\pz\pz$
with the dots representing the data and the solid histogram
the Monte Carlo simulation. The Monte Carlo simulation 
was generated using our GEANT-based
shower library. 
b) The photon shape $\chi^2$ variable for $K_L\to\pz\pz$ events.
The dots are the data and the histogram represents the Monte Carlo simulation
using our
data-based shower library. The data-based Monte Carlo simulation
shows marked improvement over the GEANT-based
shower library.
}
\end{figure}

To characterize the transverse energy deposition of a electromagnetic
shower, we devised a photon shape $\chi^2$ variable. This variable compares
the energy in the central $3\times 3$ crystals of a cluster to 
the expected energy distribution. 
While a $7\times 7$ array of crystals is used for accurate
cluster energy reconstruction, the photon shape $\chi^2$
is determined from the central $3\times 3$ crystals to
minimize any biases from accidental activity. As shown
in Figure~\ref{fig:fusecomp}, Monte Carlo events utilizing
the data-based shower library match the data better compared
to the GEANT-based shower library. For our Monte Carlo 
samples, we utilized both shower libraries.
The GEANT-based shower library was used to determine
the energy and position of the cluster, while the data-based shower
library was used for extracting the transverse shape information.

\section{Backgrounds to $K_L\to\pz\g\g$}
\label{sec:bkg}
After the event reconstruction discussed in Section~\ref{sec:recon},
large backgrounds remain in our data sample. Here we discuss
the additional criteria used to reduce these backgrounds.
The major backgrounds in our data sample consist of
events with neutral beam particles interacting in the vacuum window,
kaon decays with charged tracks, $K_L\to\pz\pz$ decays, and
$K_L\to\pz\pz\pz$ decays, with the $K_L\to\pz\pz\pz$
decays being the most difficult to remove. Vacuum window interactions can
produce $\pz\pz$ and $\pz\eta$ pairs.
To remove the vacuum window interactions, we loop over the six 
possible two photon combinations and determine the two-photon decay vertex
assuming the photons resulted from a $\pz$ decay. For each
of the six possible combinations, we reject the event
if the decay vertex is downstream of the vacuum window and the 
invariant mass of the other $\g\g$ combination is near the neutral
pion or $\eta$ mass.
Events with charged tracks are removed by requiring that the total number
of hits in the drift chamber system is less than 24; a two-track
event will produce 32 hits in the drift chambers. 

The $K_L\to\pz\pz$ events are easily identifiable because both
$\g\g$ pairs will reconstruct with $m_{\g\g} \sim m_{\pz}$ mass. Almost
all of these events are removed by rejecting events in which 
both $\g\g$ masses ($m_{12}$ and $m_{34}$) are near the
$\pz$ mass. $m_{12}$ is the two-photon invariant mass closest to the
$\pz$ mass,
while $m_{34}$ is the invariant mass of the other pair of
photons. In about two percent 
of the $K_L\to\pz\pz$ events, our choice for $m_{12}$ and
$m_{34}$ did not correctly choose both $\pz\to\g\g$ decays,
and so the cut to remove the $K_L\to\pz\pz$ background fails.
To remove these events, we also examine the other two possible
combinations of the four photons
and discard any event in which both the $m_{12}$ and $m_{34}$ 
values are near the mass of the $\pz$.

\begin{figure}
\includegraphics[width=8.5cm]{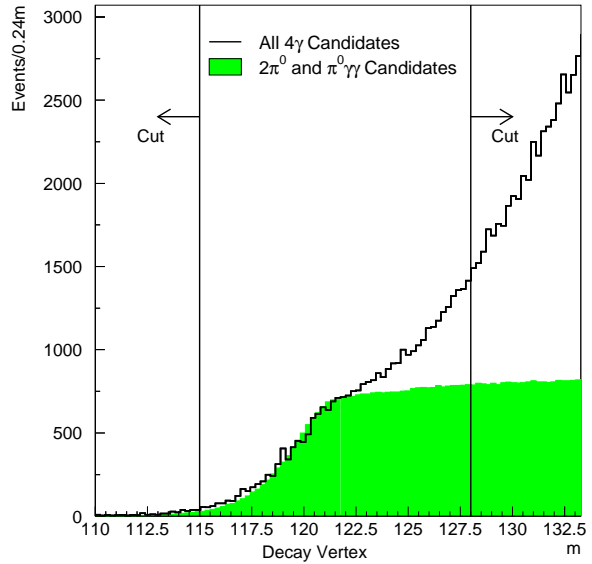}
\caption{\label{fig:zdecay} The reconstructed decay vertex position from
four photons for
all events (histogram) prior to imposing the decay vertex cut. All
cuts except for the decay vertex and photon shape $\chi^2$ have
been applied.
The shaded histogram indicates $K_L\to\pz\pz$ and $K_L\to\pz\g\g$ 
candidates.
The rise at large values is due to $K_L\to\pz\pz\pz$ decays in which
one or more photons misses the CsI calorimeter. The position of
the cut is indicated by the two vertical lines.
}
\end{figure}

Because we required exactly four photons,
$K_L\to\pz\pz\pz$ decays can only contribute to the background
if some of the photons miss the calorimeter or two or more photons
``fuse'' together in the calorimeter.
To reduce backgrounds from decays with missing photons, 
we remove events with any significant energy in any
of the photon vetoes. Also, by restricting the decay
region to $115 < z < 128$, we reduce the $K_L\to\pz\pz\pz$
background significantly because
events with missing photons tend to have a reconstructed
decay vertex downstream of
the true decay position. As shown in Figure~\ref{fig:zdecay}, the
decay vertex distribution for
$K_L\to\pz\pz$  events is relatively flat downstream of 120 meters.
However, the $K_L\to\pz\pz\pz$ background rises sharply as the
decay vertex position increases.

\begin{figure}
\includegraphics[width=8.5cm]{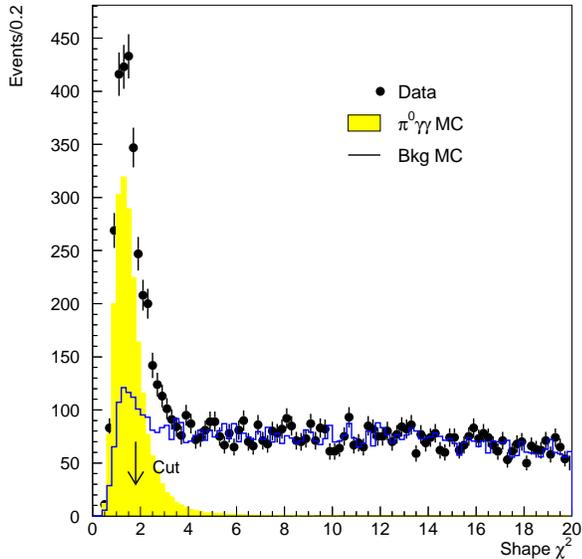}
\caption{\label{fig:fuse3x3} The photon shape $\chi^2$ variable.
The solid histogram shows the $K_L\to\pz\pz\pz$ Monte
Carlo  while the dotted histogram
is the data. The $K_L\to\pz\g\g$ signal
Monte Carlo simulation is the shaded histogram. The sum of the
signal plus background Monte Carlo simulation agrees well with the
data.
}
\end{figure}

After applying the cuts described above, there still remains a significant
number of $K_L\to\pz\pz\pz$ decays; far more than the signal from
$K_L\to\pz\g\g$. These decays result primarily from events in which 
two of the CsI clusters come from fused photons. To remove these
events, we select events with a small photon shape $\chi^2$.
For non-fused clusters this variable
peaks near zero, while for fused clusters this shape $\chi^2$ variable
becomes quite large. Figure~\ref{fig:fuse3x3} shows this variable 
for both the data and for the $K_L\to\pz\pz\pz$ background. 
We require the shape $\chi^2$ to be less than 1.8. This cut
was chosen to maximize the signal significance. For $K_L\to\pz\pz\pz$
background events we verified the 
photon shape $\chi^2$ distribution in the signal region by reweighting
events from the tails of the $\pz$ mass distribution and found the resulting
shape to correspond to our Monte Carlo prediction. The
photon shape $\chi^2$ is our final cut, and 
reduces our background to a reasonable level.

\section{Branching Ratio and $a_V$ Determination}

\begin{figure}
\includegraphics[width=8.5cm]{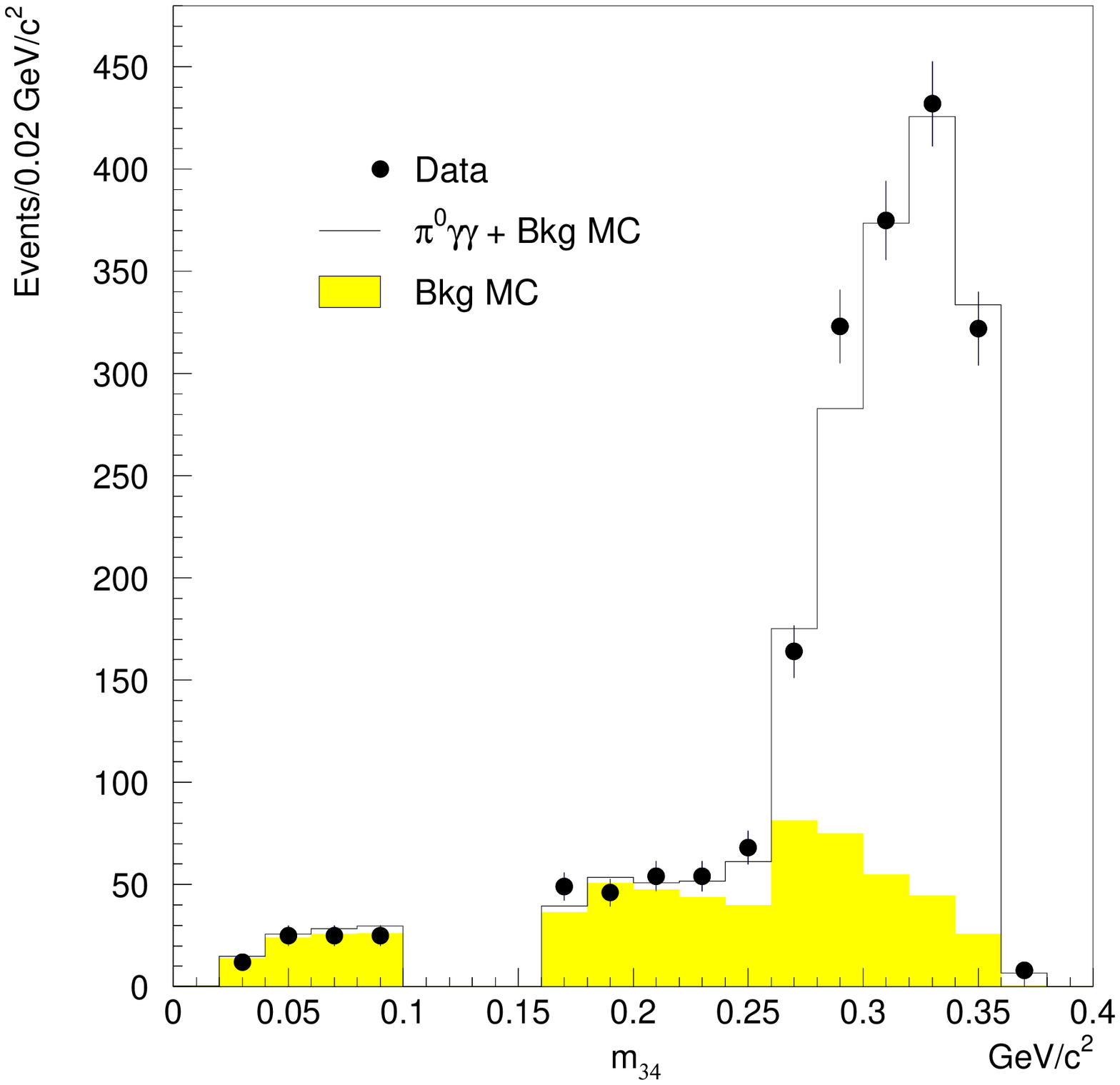}
\caption{\label{fig:zdal} 
The $m_{34}$ mass distribution for
after all selection cuts. The dots show the data, the histogram
the sum of the signal and background Monte Carlo samples, and the
shaded histogram the scaled background Monte Claro simulation.
}
\end{figure}

After applying all of the cuts described above we find 
1982 events before subtracting background.  
The final $m_{34}$ mass distribution
is shown in Figure~\ref{fig:zdal}, with the data well-represented
by the signal plus background Monte Carlo simulation. The background comprises
approximately 30\% of the total event sample.

\begin{figure}
\includegraphics[width=8.5cm]{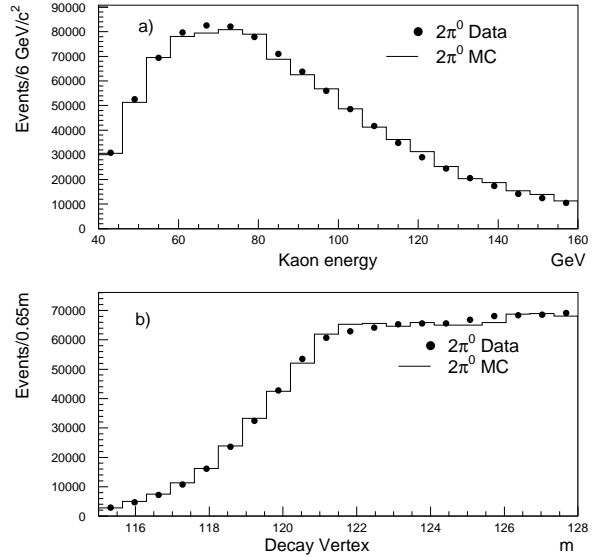}
\caption{\label{fig:2pi0_comp} The a) energy and b) decay vertex
for $K_L\to\pz\pz$ decays. The dots are the data and the
histogram is the Monte Carlo simulation.
}
\end{figure}

To determine the  $K_L\to\pz\g\g$ branching 
fraction, we use the following expression
\begin{eqnarray}
   B &=& ((N_{tot}-N_{bkg})/N_{2\pz})\times (\ep_{2\pz}/\ep_{\pz\g\g}) 
   \notag \\
      &\times& B(K_L\to\pz\pz)\times B(\pz\to\g\g),
\end{eqnarray}
where $N_{tot}$ is the number of candidate events,
$N_{bkg}$ is the number of background events, 
$N_{2\pz}$ is the number of normalization events, and
$\ep_{2\pz}$ and $\ep_{\pz\g\g}$ are the acceptances of
the $K_L\to\pz\pz$ and $K_L\to\pz\g\g$ events, respectively.
The acceptances were determined using our Monte Carlo simulation
described above.
The value $B(K_L\to\pz\pz)$ is the measured $K_L\to\pz\pz$
branching ratio. 
In the previous KTeV analysis, the value of
B($K_L\to\pz\pz$) used was $(9.36\pm 0.2)\times 10^{-4}$.
We are now using the
most recent determination of B($K_L\to\pz\pz$) 
= $(8.69\pm 0.08)\times 10^{-4}$\cite{ref:2pi0BR, ref:PDG}.
To determine the number of $K_L\to\pz\pz$
normalization decays, we count the number of events
in the region, $0.130\ \mbox{GeV/$c^2$} < m_{34} < 0.140\ \mbox{GeV/$c^2$}$.
The kaon energy and decay vertex for our normalization mode
are shown in Figure~\ref{fig:2pi0_comp}. There is good
agreement between the data and Monte Carlo simulation.

The numbers used for the branching ratio determination are shown
in Table~\ref{tab:allnumbers}. Note that the acceptances for
the signal and normalization modes are nearly identical; this
helps to significantly reduce the systematic uncertainties 
due to the acceptance calculation.

\begin{table}
  \begin{center}
  \begin{tabular}{l|rr|r} \hline\hline
   Parameter               & 1996-1997 & 1999   & Total \\ \hline
  $N_{tot}$                &   989     & 993    & 1982 \\
  $N_{tot}-N_{bkg}$        &   670.6   & 703.8  & 1374.4 \\
  $N_{2\pz}$ Events        &  482027   & 437305 & 919332 \\
  Signal Acceptance        &  0.0330   & 0.0261  & 0.030\\
  Norm   Acceptance        &  0.0328   & 0.0257  & 0.030\\
  $K_L\to 3\pz$ Bkg        &   313     & 288    & 601 \\
  $K_L\to 2\pz$ Bkg        &   5.4     & 1.2    & 6.6 \\ \hline\hline
  \end{tabular}
  \caption{Values used in branching ratio calculation}
  \label{tab:allnumbers}
  \end{center}
\end{table}

\begin{figure}
\includegraphics[width=8.5cm]{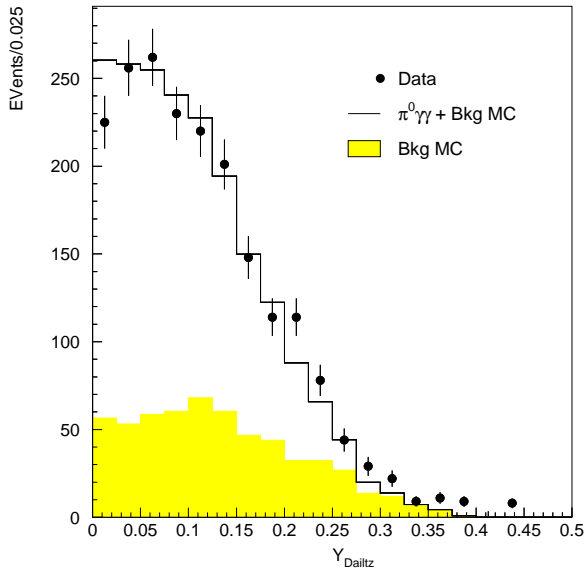}
\caption{\label{fig:ydal} 
The $Y_{Dalitz}$  distribution for
after all selection cuts. The dots show the data, the histogram
the sum of the signal and background Monte Carlo simulation, and the
shaded histogram the scaled background Monte Carlo simulation.}
\end{figure}

We also extract the value of $a_V$, using the
model described in \cite{ref:dambrosio},
from our data by performing
a two-dimensional maximum likelihood fit to the two
Dalitz parameters $Z_{Dalitz} = {m_{34}^2}/{m_K^2}$
and  $Y_{Dalitz} = ({E_{\g_3}-E_{\g_4}})/{m_K}$.
$E_{\g_3}$ and $E_{\g_4}$ are the photon energies in the kaon center-of-mass.
The distributions of the $Y_{Dalitz}$ variable is  shown in 
Figure~\ref{fig:ydal}, while the $Z_{Dalitz}$ variable is
closely related to the $m_{34}$ distribution shown in
Figure~\ref{fig:zdal}. The data used to determine 
the value of $a_V$ is listed in \cite{ref:pi0ggdata}.

\section{Systematic Uncertainties}

In general, the systematic uncertainties related to the 
branching ratio measurement are associated with either
the acceptance calculation or the background estimate.
The largest systematic error is due to the change in the acceptance
as a function of the value of $a_V$. This comes about mainly
because the $m_{34}$ distribution depends upon the  value of $a_V$,
and the acceptance varies across the $m_{34}$ region. We find
that the $a_V$ dependence has the following form:
$B = (1.33+0.13\times a_V)\times 10^{-6}$. We
conservatively evaluate this systematic error by allowing
$a_V$ to vary by $\pm 0.16$.
The next largest systematic uncertainties are related to the
acceptance ratio between the normalization and signal acceptances.
To determine a systematic error for the acceptance, we compared the
$K_L\to\pz\g\g$ data and  Monte Carlo simulation. 
We then reweighted the specific Monte Carlo distribution
to match the same distribution in data. We used this
weight factor to calculate a new acceptance and used
the difference to assign a systematic uncertainty.
We also examined the $K_L\to\pz\pz$ decays and found similar
results when reweighting the Monte Carlo simulation to match the
data. Since the $K_L\to\pz\pz$ and $K_L\to\pz\g\g$ samples
have nearly identical acceptances, this gives us confidence 
in our estimate in the systematic effects.
The variables that had the largest effect on the acceptance
were the kaon energy and the photon veto response.
We assigned a systematic uncertainty of 1.16\% due to the acceptance.

The ability of our Monte Carlo simulations to reproduce the $K_L\to\pz\pz\pz$ 
background also contributes to the systematic uncertainty. To estimate the effects
from our knowledge of the background, we looked at all events before
applying the shape $\chi^2$ cut. This sample is dominated by 
$K_L\to\pz\pz\pz$ events. We then reweighted the background Monte Carlo sample to match
the data in a particular parameter. 
The change in acceptance multiplied by the background fraction
was taken to be the systematic uncertainty from a specific variable.
In particular, we assigned the following systematic uncertainties
due to our simulation of the background: the photon shape $\chi^2$ (1.07\%),
the drift chamber simulation (0.92\%), the photon veto simulation (0.90\%), 
the kaon energy shape (0.69\%), and the
$K_L$ decay vertex distribution (0.38\%). 

In addition to the acceptance calculation and the background determination,
a few other effects contribute to our systematic uncertainty including
the Monte Carlo statistics, the background normalization and the 
measured $K_L\to\pz\pz$ and $\pz\to\g\g$ branching fractions.
For this analysis, we generated nearly
$10^{11}$ $K_L\to\pz\pz\pz$ decays, more than twice the
background statistics. These
statistics contribute 1.0\% to the systematic uncertainty.
To determine the normalization of the $K_L\to\pz\pz\pz$ Monte Carlo
sample, we first scaled the
$K_L\to\pz\pz$ Monte Carlo sample 
to the observed $\pz$ peak in the $m_{34}$ mass
distribution in the data. 
We then normalized the $K_L\to\pz\pz\pz$ sample relative
to the number of $K_L\to\pz\pz$ events by the ratio of the
branching ratios and the number of generated events. To assign
a background normalization systematic error, we
scaled the $K_L\to\pz\pz\pz$ background events directly  to the
shape $\chi^2$ distribution and compared the difference between
the two methods. This contribututed  0.90\% to the total
systematic uncertainty.  Finally, we assigned a 0.5\% systematic
uncertaintly due to  the error on the measured
$K_L\to\pz\pz$ branching ratio.
All of the systematic effects are listed in Table~\ref{tab:systematics},
with a total systematic uncertainty on the $K_L\to\pz\g\g$
branching ratio of 3.0\%.

\begin{table}[htbp]
\centerline{ 
  \begin{tabular}{llc} \hline\hline
  Type   & Source                    & Uncertainty (\%) \\ \hline
  Acceptance &  $a_V$ dependence          & 1.50 \\ 
             &  MC acceptance ratio       & 1.16 \\ \hline
  Background &  Photon shape $\chi^2$     & 1.07 \\
             &  Drift chamber hits        & 0.92 \\
             &  Photon vetoes             & 0.90 \\
             &  Kaon energy               & 0.69 \\
             &  Decay Vertex              & 0.38 \\ \hline
  General    &  MC statistics             & 1.00 \\
             &  Background normalization  & 0.90 \\
             &  $K_L\to 2\pz$ branching ratio & 0.50 \\ \hline
             &  Total                     & 3.0 \\  \hline \hline
  \end{tabular}
}
\caption{Branching ratio systematic uncertainties}
\label{tab:systematics}
\end{table}

To determine the systematic uncertainty in our $a_V$ measurement,
we varied the position of the selection cuts and looked for
any non-statistical change in the value of $a_V$. We also 
varied the level of $K_L\to\pz\pz\pz$ background 
according to the methods described above.
The
major systematic uncertainties associated with the determination of
$a_V$ are listed in Table~\ref{tab:avsyserr}. The main sources
of systematic error result from the uncertainty of the
background estimations.
The total
systematic uncertainty associated with the $a_V$ measurement
is 0.07.

\begin{table}[htbp]
\begin{center}
\begin{tabular}{lc} \hline\hline
Source                 & Uncertainty  \\\hline
Z  vertex cut          & 0.05 \\ 
Photon veto cut        & 0.04 \\ 
3$\pz$ normalization   & 0.03 \\ 
Photon shape $\chi^2$  & 0.01  \\  \hline
Total  & 0.07  \\ \hline\hline
\end{tabular}
\end{center}
\caption{$a_V$ fitting systematic uncertainty.}
\label{tab:avsyserr}
\end{table}

\section{Final Results and Conclusions}
To obtain the final branching ratio result, 
we used the weighted average of the
1996-1997 and 1999 numbers based upon the statistical
errors of the two results.
The systematic studies were done on the combined 
1997 and 1999 analyses to take into account any correlations.
Including the uncertainties due to the
systematic effects, we find the $K_L\to\pz\g\g$ branching ratio to be
\begin{eqnarray}
B(K_L&\to&\pz\g\g) =\notag \\
&&(1.29\pm0.03_{stat}\pm0.05_{syst})\times 10^{-6}.
\end{eqnarray}
This result is a significant improvement over the previous KTeV
result, and supersedes that result. The differences between the
current and previous results is discussed in Section~\ref{sec:appA}.

Our value of $a_V$ was obtained using the fitting method
described above. 
The $\chi^2$ for the fit is 
56.6 for 59 degrees of freedom.
Including the systematic error, we find
\begin{eqnarray}
a_V = -0.31\pm 0.05_{stat}\pm 0.07_{syst}.
\end{eqnarray}
The total error from our determination of $a_V$ is slightly
larger than the NA48 result, however, it is compatible with
their value.

The branching ratio result is consistent with the latest
$O(p^6)$ ChPT results. Our value of $a_V$ suggests that
the CP conserving amplitude in $K_L\to\pz\ll$ should be
less than $1\times 10^{-12}$ compared to the expected
total branching ratio of $\sim 3\times 10^{-11}$\cite{ref:buchalla}. 
Therefore this decay should be dominated by
CP violating terms. Future searches for $K_L\to\pz\ee$
and $K_L\to\pz\mm$
would be of great interest since many models of  new physics would 
signficantly alter these branching ratios.

\section{Appendix A}
\label{sec:appA}
Compared to the previous KTeV  $K_L\to\pz\g\g$ branching ratio
value, our
new result is significantly lower. The main difference between
the two analyses arises from our simulation of the transverse
photon shower shape. Our previous analysis used the GEANT-based
shower library, while our current analysis utilizes the
data-based shower library. As shown in 
Figure~\ref{fig:fusecomp}, the data-based shower library
shows significant improvement over the GEANT-based
shower library.
Utilizing this new
shower library changes our estimate of the $K_L\to\pz\pz\pz$
background, increasing the background by a factor of approximately two. 
The increase in background occurs because the $K_L\to\pz\pz\pz$ background 
peaks in the region of small shape $\chi^2$ as shown in 
Figure~\ref{fig:fuse3x3}. In our previous result
the $K_L\to\pz\pz\pz$ background shape dropped
in the signal region. Accordingly, the background
estimate utilizing the GEANT-based shower library
underestimated the $K_L\to\pz\pz\pz$ background.

In the previous result the systematic error for the mismatch
in the photon shape $\chi^2$ scaled with the size of the estimated background.
However,
since the background shape was incorrectly modeled, our estimate of the
systematic error also was underestimated.
Studies of other variables sensitive to
the $K_L\to\pz\pz\pz$ background arrive at a similar estimate
for the background level in the current analysis.

\begin{acknowledgments}

We gratefully acknowledge the support and effort of the Fermilab
staff and the technical staffs of the participating institutions for
their vital contributions.  This work was supported in part by the U.S.
Department of Energy, The National Science Foundation, The Ministry of
Education and Science of Japan,
Fundação de Amparo a Pesquisa do Estado de S\~ao Paulo-FAPESP,
Conselho Nacional de Desenvolvimento Cientifico e Tecnologico-CNPq and
CAPES-Ministerio Educao.
\end{acknowledgments}


\end{document}